## RESEARCH ARTICLE

# A Framework for Migrating to Post-Quantum Cryptography: Security Dependency Analysis and Case Studies

**KHONDOKAR FIDA HASAN**[1], **(Senior Member, IEEE), LEONIE SIMPSON**[2],
**MIR ALI REZAZADEH BAEE**[2], **(Senior Member, IEEE), CHADNI ISLAM**[2], **ZIAUR RAHMAN**[2],
**WARREN ARMSTRONG**[3], **PRAVEEN GAURAVARAM**[4], **AND MATTHEW MCKAGUE**[2]

[1]Canberra School of Professional Studies, University of New South Wales (UNSW), Canberra, ACT 2601, Australia
[2]School of Computer Science, Queensland University of Technology (QUT), Brisbane, QLD 4000, Australia
[3]QuintessenceLabs Pty Ltd., Canberra, ACT 2609, Australia
[4]Cybersecurity Research & Innovation, Tata Consultancy Services Ltd., Australia & New Zealand, Brisbane, QLD 4000, Australia

Corresponding author: Matthew McKague (matthew.mckague@qut.edu.au)

The work was supported by the Cyber Security Research Centre Ltd. funded by the Australian Government's Cooperative Research Centers Program.

**ABSTRACT** Quantum computing is emerging as a significant threat to information protected by widely used cryptographic systems. Cryptographic methods, once deemed secure for decades, are now at risk of being compromised, posing a massive threat to the security of sensitive data and communications across enterprises worldwide. As a result, there is an urgent need to migrate to quantum-resistant cryptographic systems. This is no simple task. Migrating to a quantum-safe state is a complex process, and many organisations lack the in-house expertise to navigate this transition without guidance. In this paper, we present a comprehensive framework designed to assist enterprises with this migration. Our framework outlines essential steps involved in the cryptographic migration process, and leverages existing organisational inventories. The framework facilitates the efficient identification of cryptographic assets and can be integrated with other enterprise frameworks smoothly. To underscore its practicality and effectiveness, we have incorporated case studies that utilise graph-theoretic techniques to pinpoint and assess cryptographic dependencies. This is useful in prioritising crypto-systems for replacement.

## I. INTRODUCTION

Cryptography underpins the security resiliency of organisations in today's digital landscape. Fundamentally, cryptography is about the application of mathematical algorithms to provide security services, such as confidentiality, integrity, authenticity, or non-repudiation of data [1]. These security services are illustrated in FIGURE 1 [2], [3].

Cryptographic systems are often deployed to protect high-value data, and so are often a target for attack. The term *cryptanalysis* refers to the analysis of a cryptographic system

The associate editor coordinating the review of this manuscript and approving it for publication was Wei Huang.

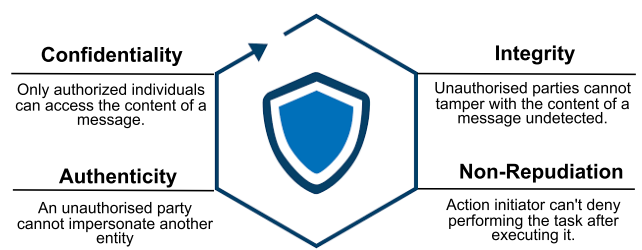

**FIGURE 1.** Foundational security services.

in an effort to discover weaknesses [4], [5]. These weaknesses can be at the algorithmic level (for example, in the design of the cryptographic algorithms used), at the parameter level (for example, specifying insufficient key lengths for the length of protection required) or at the implementation level (for example, having timing or power-analysis side-channels in the algorithm's implementation). Any key has a finite size and so, in principle, it is possible to attack a system by trying all possible key values. This approach is termed a *brute-force attack*. For a secure algorithm, it must be infeasible to determine the correct key within a reasonable time (for example, requiring centuries of trial-and-error) [6]. The defence against brute force attacks on the key is to increase the key length.

There are two main categories of cryptographic techniques: symmetric and asymmetric. Some cryptographic techniques are based on the difficulty of solving an underlying complex mathematical problem. For example, the asymmetric cryptographic algorithm RSA relies on the difficulty of factoring a very large number that is the product of two large primes. Another cryptosystem, Elliptic Curve Digital Signature Algorithm (ECDSA), relies on solving the discrete logarithm problem [1]. These two systems are fundamental building blocks for many of today's widely-used cryptosystems, and protocols such as Transport Layer Security (TLS) used for HTTPS traffic. Similarly, in symmetric cryptography, the Advanced Encryption Standard (AES) is a widely accepted block cipher algorithm. Note that AES does not rely on the difficulty of solving a mathematical problem in the same sense as RSA or ECDSA do. The strength of AES comes from its complex series of substitutions, permutations, and transformations [1], [7], [8].

The emergence of quantum computing presents a new avenue for cryptanalysis. Quantum computers can significantly reduce the complexity of key search operations, reducing the security level of symmetric cryptosystems. Additionally, quantum computers offer efficient solutions to factoring and discrete logarithm problems, which form the foundation of many widely deployed asymmetric cryptosystems. As a consequence, quantum computers could potentially reduce the time required to compromise certain well-established cryptosystems from durations unfeasible with classical computers (e.g., billions of billions of years) to mere days or even less. This poses a significant challenge to the security guarantees of many traditional cryptographic techniques [9]. Hence, the advent of quantum computing has put some of the widely used traditional cryptography, especially public key cryptography, under threat, as it has the potential to break many of the popularly deployed cryptographic algorithms, including RSA.

Some recently proposed cryptosystems are designed to be resistant to quantum attacks. These are collectively known as Post Quantum Cryptography (PQC) [10]. The National Institute of Standards and Technology (NIST) has initiated steps to standardise PQC primitives implementing at least one of the functionalities of public key encryption, key encapsulation mechanism (KEM), or digital signature. This stadardisation process is nearing the finalisation stage [11].

Some symmetric cryptosystems, such as AES [12] are thought to be secure against quantum attacks when used with larger keys.

### A. RESEARCH MOTIVATION

It is critical for organisations to migrate insecure cryptosystems to a quantum-safe state in a timely manner [13]. However, the enterprise migration to quantum-safe cryptography is unlikely to be straightforward due to a variety of factors:

- PQC algorithms have different resource requirements from traditional asymmetric cryptosystems; including having much longer keys. This may require updates to protocols and key management
- Many hardware appliances that include hardware acceleration for widely used cryptosystems must be replaced
- Legacy systems without continuing software support may need to be replaced
- Compatibility with hybrid and legacy systems
- Maintaining regulatory compliance
- Many organisations lack staff with relevant expertise

These factors mean that such migration will be costly and time-consuming [14], [15], [16].

For a resilient enterprise system, the necessity of migration to a quantum-safe state is undeniable. However, the exact cost and time required for such migration is unknown. Adopting a systematic approach to this transition helps organisations migrate smoothly and reduces the chance of introducing new vulnerabilities [14], [16]. Hence, a robust framework for PQC migration is required, encompassing inventories and security classification of assets and processes, determining priorities, risk assessment, implementation planning, monitoring and review.

As part of determining migration priorities, the framework should identify the most vulnerable information assets to migrate to a quantum-safe state first. Hence, the framework must be able to identify which cryptosystems impact the security of each information asset. For large organisations this can be a complex task, which can be broken down into two main parts:

- Identify sources of information about information assets, process, and their relationships.
- Use this information to trace which cryptographic primitives affect each data asset, organised by the level of data classification for the data.

### B. RESEARCH CONTRIBUTION

In this paper, we propose a framework for PQC migration that draws on existing standards and practices for information asset and cryptographic asset (keys, certificates, etc.) management in enterprise settings. By leveraging these, our framework provides a step-by-step method to use existing inventories of data, IT assets, and cryptographic assets and to identify dependencies between data assets and cryptosystems. The framework allows for effective re-evaluation upon changes to organisation inventories or the security status of cryptosystems. The robustness of the framework stems from its ability to respond to changes in both information assets and the cryptographic landscape.

There are two main contributions in this paper:

1) A high level framework to guide PQC migration for enterprise systems
2) A detailed security dependency analysis process that identifies situations where cryptosystems in use do not provide the security level desired for the information asset over the desired time frame

We demonstrate the utility of the dependency analysis process through three case studies. Furthermore, this paper provides valuable insights into the quantum threat, associated risk, and actionable steps, including inventory compilations, as well as practical advice for developing a plan for migrating an organisation to PQC, which can be useful for both academics and practitioners.

Our framework is flexible and can be used in conjunction with other security practices and frameworks. In particular, our security dependency analysis provides a new component that is lacking in other frameworks.

### C. ORGANISATION OF THE PAPER

At a high level, this paper can be seen as a three-step development as shown in FIGURE 2. The remaining sections of this paper can be found as follows.

(1.)
   **Section II** presents quantum computing and its potential impact on traditional cryptographic systems, motivating the need for migrating to new cryptographic solutions.
   **Section III** delves into the topic of post-quantum cryptography, describing its key concepts, and methods for protecting information from quantum attacks along with the timeline for migration.

(2.)
   **Section IV** explores the critical issues of enterprise information asset security and data classifications.
   **Section V** presents a discussion on cryptographic inventories, their importance, challenges, requirements, and current industry practices for constructing them.
   **Section VI** focusses on the current state of the art in frameworks for PQC migration.

(3.)
   **Section VII** presents the proposed framework for PQC migration based on security dependencies.
   **Section VIII** gives the general rule of thumb to evaluate security dependencies, followed by three case studies in **Section VIII-B**, **Section VIII-C** and **Section VIII-D** before the conclusions of the paper in **Section IX**.

For the convenience of the reader, a list of abbreviations used throughout the paper is given in Table 1.

## II. QUANTUM THREAT AND RISK OVERVIEW

This section gives an overview of quantum computing and outlines quantum threats to classical cryptography.

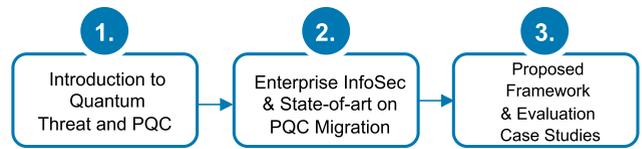

**FIGURE 2. High-level structure of the paper.**

**TABLE 1. List of acronyms used in the paper.**

| Acronym | Meaning |
|---------|---------|
| ACSC | Australian Cyber Security Centre |
| AES | Advanced Encryption Standard |
| CARAF | Crypto Agility Risk Assessment Framework |
| DL | Discrete Logarithm |
| EDR | Endpoint Detection and Response |
| EFS | Encrypted File System |
| ETSI | European Telecommunications Standards Institute |
| IAR | Information Asset Register |
| ECDSA | Elliptic Curve Digital Signature Algorithm |
| EIS | Enterprise Information Security |
| GDPR | General Data Protection Regulation |
| HIPAA | Health Insurance Portability & Accountability Act |
| ISG | Industry Specification Group |
| PQC | Post Quantum Cryptography |
| PSPF | Protective Security Policy Framework |
| QSC | Quantum-Safe Cryptography |
| NIST | National Institute of Standards and Technology |
| NIST CSF | NIST Cyber Security Framework |
| NISTIR | NIST Internal or Interagency Report |
| NCSA | National Cyber Security Centre |
| RACI | Responsible, Accountable, Consulted and Informed |
| RSA | Rivest-Shamir-Adleman |
| SIEM | Security information and event management |
| SSL | Secure Sockets Layer |
| SOX | Sarbanes-Oxley Act |
| TLS | Transport Layer Security |

### A. OVERVIEW OF QUANTUM COMPUTING

Quantum computing refers to a class of technologies that aim to take advantage of quantum effects, such as superposition, in order to speed up computation. Superposition enables the implicit calculation of multiple values at once. However, getting meaningful data out of such a superposition is non-trivial, and only a few classes of algorithms have been found that can solve problems faster than classical (i.e., non-quantum) computers [17]. There are three general classes of problems where quantum computers can outperform classical computers: search, hidden subgroups, and quantum simulation.

Search problems involve trying all possible answers and determining the correct one. Grover's algorithm [18] is a quantum algorithm for the search problem. While a classical computer needs time proportional to $n$ to search through $n$ possible answers, Grover's algorithm solves the same problem in time proportional to the square root of $n$. For example, if $n$ is 1 million, then Grover's algorithm is potentially about 1,000 times faster than a classical computer. This advantage scales even further as the value of $n$ increases. Search is important in many applications, including optimisation.

The most familiar example of the hidden subgroup problem is factoring, where we are given a large number that is the product of two large prime numbers, and we seek to find out what those prime numbers are. Classically, this is a difficult problem, but it can be solved quickly on a quantum computer using Shor's algorithm [19].

Finally, as a quantum device, a quantum computer can directly simulate the quantum effects of physical systems [20]. This is a great advantage in both speed and memory over classical simulations, which require an exponential amount of memory and time, limiting them to small simulations only.

Quantum computers, once developed, will likely see applications in science related to quantum simulation and optimisation, as well as in businesses where optimisation is important. They are unlikely to be useful for typical personal computer applications, at least for the foreseeable future.

The physical quantum effects required to build quantum computers are extremely fragile. For computation to be successful, the quantum computer needs to be extremely well isolated while simultaneously allowing for interaction with the outside world for control and readout. These conditions are difficult to meet at scale. To date, the largest quantum computer announced has 433 qubits [21]. Although some computations, particularly quantum simulations, can be done with a small number of qubits, many applications, including cryptanalysis, will require millions of qubits. While there has been steady progress, quantum computers of this scale are likely decades away [22].

### B. QUANTUM THREAT TO CRYPTOGRAPHY

Quantum computing has an impact on the security provided by cryptosystems. Modern cryptography, aside from specialised information theoretic protocols, depends on hard computational problems for their security. In particular, they depend on hard problems where there is an easy solution if you know some special information.

Symmetric cryptosystems depend on the difficulty of searching through a large number of possible secret values. Advanced Encryption Standard (AES) [12] is one such system where exhaustive key search is the greatest threat. Quantum computers, using Grover's algorithm, can find the key used with AES in much less time than a classical computer.

Asymmetric cryptosystems, meanwhile, depend on mathematical problems rather than search. An important example is RSA [23], which depends on the difficulty of factoring. To encrypt a message, one needs to know a large number $n$, which is the product of two large primes $p$ and $q$. To decrypt a message, one needs to know $p$ and $q$. The security of RSA depends on the fact that it is difficult to find $p$ and $q$ given $n$. Quantum computers break this assumption because they can find $p$ and $q$ very easily using Shor's algorithm. Many other widely used cryptosystems depend on similar assumptions, such as the hardness of finding discrete logarithms (DL), including Diffie-Hellman, El Gammal, and elliptic curve cryptography [24]. Due to the speed of Shor's and similar algorithms, these cryptosystems have almost no security against adversaries with quantum computing capabilities.

TABLE 2 shows an estimation of the quantum resources required to break some popular cryptosystems. For comparison, we selected Discrete Logarithm (DL) over the NIST P-256 elliptic curve [26], RSA with 3072-bit modulus, and AES with a 128-bit key. All of these cryptosystems are considered to have 128 bits of security against classical adversaries [25]. Note that the resources (size and time) required to break these cryptosystems are not the same. This has implications for how quickly these cryptosystems become vulnerable to attacks as quantum computers grow in size.

For symmetric cryptosystems, Grover's search does not have as much advantage over classical algorithms as Shor's algorithm. Hence breaking AES on a quantum computer is estimated to be extremely time-consuming and expensive compared to attacking vulnerable asymmetric cryptosystems with quantum computers. For example, according to one model [25], to break AES-128 requires a quantum computer 22 orders of magnitude larger than for breaking RSA. Further, its estimated, such a computer will take one year to break AES-128 compared to only one-day to break an asymmetric algorithm. The overall security level for AES-128 in this model is about 106, rather than 128 against classical computers. Because of this difference, symmetric cryptosystems like AES will retain a relatively high level of resilience against quantum adversaries, even when quantum computers of this size exist.

Among asymmetric cryptosystems, there are smaller but still significant differences between the quantum resources required to break different cryptosystems. Elliptic curve cryptosystems [24], [27], currently popular because of their efficiency, are the most vulnerable. Breaking RSA, on the other hand, requires about 10 times the quantum memory of elliptic curve cryptosystems, potentially extending its usefulness by several years.

The difference noted above means that, *it is important to properly prioritise the replacement of each type of cryptosystem*. Other considerations (data security classifications, time required for data to remain secure) being equal, *elliptic curve cryptosystems should be replaced first*.

When comparing classical and quantum attacks on cryptosystems, it is important to consider some important facts:

- **Quantum computers already pose a significant risk.** Expert opinion is that there is about 1% to 6% chance that cryptographically significant quantum computers will be built in the next 5 years [22]. While that may seem low, it is extremely likely compared to breaking a mature cryptosystem with classical computing using currently known methods. Hence quantum computers already need to be treated as a credible threat.

**TABLE 2.** Estimates of quantum resources required to break popular cryptosystems. Each listed algorithm is equivalent to 128-bits of security against classical adversaries [25].

| Algorithm | Size of Quantum Computer | Time Required |
|---|---|---|
| DL with NIST P-256 | $6.8 \times 10^7$ qubits | 24 hours |
| RSA-3072 | $6.4 \times 10^8$ qubits | 24 hours |
| AES-128 | $10^{30}$ qubits | 1 year |

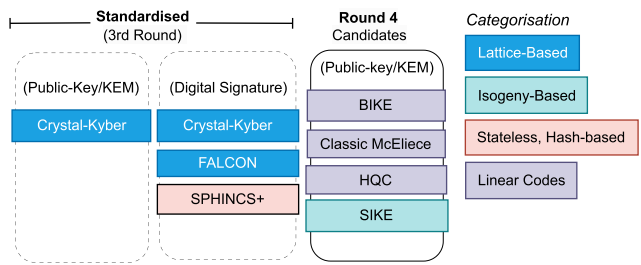

**FIGURE 3.** NIST final standardised PQC after 3rd round of the competition and round 4 candidates [28], [29].

- **Many data assets need to be protected for extended time frames.** While it may take a long time to build quantum computers large enough to threaten cryptosystems, we must also consider the time required to deploy quantum-safe alternatives and the length of time that data must be kept secure. Encrypted data stored now may become vulnerable during its lifetime, an attack strategy known as "store now, decrypt later".

- **A quantum computer is a one-time cost for an attacker.** For classical attacks against asymmetric cryptosystems, attacking each new key is extremely difficult. For quantum attacks, the difficulty is in building the quantum computer, after which any key for a vulnerable public-key cryptosystem can be attacked relatively easily. This makes quantum attacks a much better value proposition to an adversary with abundant resources to spend.

## III. QUANTUM-SAFE CRYPTOGRAPHY AND MIGRATION TIMELINE
This section provides the context and overview of NIST PQC endeavour, threats to it and standard timeline for migration.

### A. POST-QUANTUM CRYPTOGRAPHY
PQC refers to the branch of cryptography involved in developing cryptosystems and protocols resistant to quantum attacks. PQC attempts to create new cryptosystems based on mathematical problems that are thought to be difficult even for quantum computers. Examples of post-quantum cryptographic schemes include lattice-based cryptography, code-based cryptography, hash-based cryptography, and multivariate cryptography [11], [14]. With the expectation that quantum computers will become more powerful and accessible in the near future, PQC will be a necessary tool to ensure the security of the digital ecosystem.

### B. QUANTUM-SAFE CRYPTOGRAPHIC ALGORITHMS
The National Institute of Standards and Technology (NIST) is leading an effort to develop and standardise post-quantum cryptographic algorithms resistant to quantum attacks [12]. The NIST PQC standardisation project, launched in 2016, is a collaborative process involving the cryptographic community in identifying and evaluating new post-quantum cryptographic algorithms. The goal of the project is to provide a set of standardised, interoperable, and quantum-safe cryptosystems that can be used by government agencies, businesses, and other organisations to protect sensitive information from future quantum attacks.

The NIST project has involved multiple rounds of public review and evaluation, and the final set of standardised post-quantum cryptographic algorithms is expected to be released in the coming years. After the 3rd round, NIST has identified the following algorithms for standardisation [12]:

1) CRYSTALS-KYBER (key-establishment) and CRYSTALS-Dilithium (digital signature) are lattice-based algorithms. They were selected for their robust security and excellent performance, and NIST expects them to work well in most applications.

2) FALCON (digital signature) is another lattice-based algorithm that may be used in cases for which CRYSTALS-Dilithium signatures are too large.

3) SPHINCS+ (digital signature) is based on hash functions. It will also be standardised to avoid relying only on the security of lattices for signatures.

Round 4 of the NIST project is ongoing at the time of writing and involves four other cryptographic algorithms (BIKE, Classic McEliece, HQC, and SIKE [28]), as detailed in FIGURE 3.

### C. THREATS TO POST-QUANTUM CRYPTOGRAPHY
The previous section discussed quantum threats to classical cryptosystems. However, threats to post-quantum cryptography also exist. The quantum threat to PQC is based not only on the theoretical possibility of a large-scale quantum computer but also on practical advances in quantum or classical algorithms capable of attacking specific PQC schemes. For example, some quantum algorithms have been proposed to break the security of code-based, multivariate, and isogeny-based PQC schemes [30], [31].

The NIST PQC standardization process has flagged some candidate algorithms that were found to be vulnerable to quantum attacks, such as BIKE, HQC, and SIKE [32], [33]. At the same time, the remaining candidates are still subject to further scrutiny and analysis, and may face new quantum or classical threats in the future [34]. Because most PQC cryptosystems are relatively new and less studied than older cryptosystems, these kinds of advances are more likely.

As a result, it is clear that the quantum threat to PQC is a dynamic and evolving factor that must be constantly monitored and evaluated. The migration should include strategies for dealing with possible compromises of PQC cryptosystems, such as updating the parameters, algorithms, or protocols of PQC schemes or switching to more resistance to quantum attacks PQC schemes.

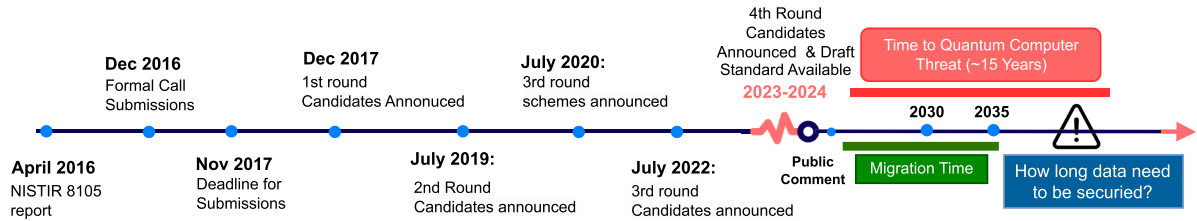

**FIGURE 4.** Projected Timeline for PQC migration [22].

Another strategy to mitigate threats to PQC is to adopt hybrid cryptosystems (see eg. [35], [36], [37]) which combine a PQC cryptosystem with a well-studied cryptosystem like RSA. Hybrid systems offer the best of both worlds as their security depends on the *stronger* cryptosystem, meaning that even if the PQC primitive is classically compromised then the security is still as good traditional cryptosystem. Hybrid systems are also useful for migration as they can often be implemented as modifications to existing protocols like x.509 or TLS [35], [37], offering backwards compatibility for legacy systems that have yet to be migrated.

### D. MIGRATION TO PQC AND STANDARD TIMELINE

Migration to PQC is required to ensure the security of communication and data transmission [14], [16], [38]. This involves transitioning from traditional cryptographic methods to post-quantum cryptographic algorithms that are resistant to quantum attacks. The migration process includes identifying systems and applications for an upgrade, selecting and integrating suitable post-quantum cryptographic algorithms into infrastructure, and validating the security of these new methods through rigorous testing to ensure resilience against quantum threats. The PQC migration process can be complex and time-consuming. It necessitates careful planning and coordination to ensure that the migration is carried out correctly and without disrupting the services provided by the systems and applications.

FIGURE 4 shows a timeline indicating the PQC standardisation and migration processes. The depicted hypothetical migration timeline, represented by the green line, suggests that the transition to PQC should ideally occur within the next 15 years, with enterprise systems transitioning by approximately 2035. However, this timeline is not universally applicable because risk management strategies must be consistent with data characteristics and longevity requirements. More specifically, the duration for which data must be secured, shown in the blue line, is critical in determining the urgency and scale of the migration. Therefore, the urgency and scale of this migration will, in fact, vary depending on how long the data must be protected [15], [16]. Furthermore, as mentioned in Section II-B, the threat quantum computers pose depends on the type of cryptosystem in use.

Since April 2016, NIST has been supervising the PQC migration process. Beginning with the NISTIR 8105 reporting, four rounds of global competition were used to identify and fine-tune the selection of PQC primitives. At the time of writing in mid 2023 NIST has not yet announced the final candidates [14]. As 2024 approaches, early adopters are expected to begin integrating the standardised quantum-resistant algorithms into their systems, bringing in a new era of quantum-resistant cybersecurity.

## IV. BACKGROUND OF ENTERPRISE INFORMATION SECURITY

Enterprise Information Security (EIS), in general, refers to the prevention of unauthorised access, use, disclosure, disruption, modification, or destruction of an organisation's information and information systems. EIS also refers to various activities and technologies that protect an organisation's digital infrastructure, data, and intellectual property. This section briefly presents some concepts, processes, and practises used to enable enterprise information security, such as the relevant concepts of Information Asset, Asset Register, and Information Asset and Security Classification Procedure. Our framework will operate within the context of these established practices and make use of them as part of the migration process.

### A. INFORMATION ASSETS

An information asset can be defined as a piece of information that is valuable to an organisation and requires protection from unauthorised access, modification, or disclosure. Information assets can take many forms, including electronic files, physical documents, intellectual property, customer data, and financial records [39]. For example, a company's customer database contains sensitive information like names, addresses, and payment details, which are information assets. Similarly, a company's trade secrets or proprietary research could be considered information assets, as they provide a competitive advantage and must be protected from unauthorised access, modification or disclosure. NIST recommends that organisations classify their information assets based on their value and sensitivity to ensure that appropriate security controls are in place to protect them.

Overall, information assets are considered a critical component of enterprises' overall security systems, and their protection is a key priority since most organisations rely on at least some sensitive information to operate.

### a: TERMINOLOGY: ASSET REGISTER, INFORMATION ASSET REGISTER, ASSET INVENTORY AND DATA INVENTORY

Generally, an *Asset Register* refers to a list or database of an organisation's assets, including physical equipment (i.e., hardware), software, and other resources. Such an asset register typically includes information about each asset, which may include its location, purchase date, cost, condition, maintenance schedule, and other relevant details [40], [41]. Asset registers are considered an essential tool for managing assets in an organisation. They are used for various purposes, such as tracking the location and status of assets, monitoring their usage, and planning for maintenance and replacement. They can also be used to assess the overall value of an organisation's assets and ensure that they are used efficiently and effectively [40], [42].

In the context of enterprise information security, an asset register, referred to as the *Information Asset Register (IAR)*, can be used to track an organisation's information systems, data assets and physical assets. An IAR is a structured and comprehensive list of information assets and associated metadata (e.g., ownership, classification, retention requirements, and access restrictions) that an organisation uses to manage its information assets throughout its lifecycle [39]. It is accessed and managed to support business needs and to ensure effective asset control. Organisations can avoid unnecessary or unintentional duplication of information, systems, and processes by maintaining an IAR [43]. By maintaining an accurate and up-to-date IAR, organisations establish security policies and controls to protect assets, monitor and manage their use and access, improve their overall security posture and reduce the risk of security breaches [41].

However, as the practitioner guide of the office of the Victorian Information Commissioner states [39], there is no single set process or a way for an organisation to define its information assets because the definition should fit the business needs of the organisation. In practice, organisations should describe their information assets in enough detail to make it possible to manage each of them and the whole. If the registry is not comprehensive enough, the organisation might not have enough information to properly manage assets. If it is too detailed, the organisation will waste resources on gathering and maintaining unused information. The attributes or metadata used to describe information assets vary by organisation. These attributes should reveal the individual characteristics or features of information items, which can then be combined into a more comprehensive form that is useful to the organisation. This comprehensive form can be based on specific collections, functions, subjects, or processes and is considered an information asset unique to that organisation [39].

In a nutshell, although these terms are distinctive, an organisation may use one or more of the following to manage their assets, and often they use them as an interchangeable asset in application [43]:

- An Asset Register, in general, which is a comprehensive record of all fixed assets owned by an organisation, detailing their purchase date, cost, value, and other relevant information.
- An Inventory[1] or Asset Inventory, in general, which is a detailed list of all goods or materials that a business holds, either for the purpose of resale or for use in production, including their quantities and values.
- A Catalogue, in general, which originally indicates a descriptive list or record of items, often products or services may be offered by a business, presented with descriptions, prices, and sometimes images for customer reference.

### B. INFORMATION ASSETS AND SECURITY CLASSIFICATION

In general, organisations classify information to help users understand the security requirements associated with handling various types of data. Classification policies describe information security levels and methods for assigning information to specific classification levels. The storage, handling, and access requirements for classified data are determined by the security categories or classifications used by an organisation. These classifications are assigned based on the information's sensitivity and its importance to the organisation.

Although classification systems differ, they essentially aim to separate public and private information and classify data into high, medium, and low sensitivity levels. For instance, the government uses terms such as top secret, secret, protected, official: sensitive, and official, whereas a company might use terms like highly sensitive, sensitive, internal, and public to achieve the same goal [44], [45].

Data classification is crucial because it serves as the foundation for data security decisions. For example, to protect sensitive and highly sensitive information both at rest, in use, and in transit, a business might require the use of strong encryption. This illustrates a requirement for data handling [44] that needs to be translated into policies and procedures. Organisations implement labelling specifications that apply consistent markings to sensitive information when classifying information. Using standard labelling practices, users will always be able to recognise sensitive information and handle it properly [44].

### a: INFORMATION ASSETS SECURITY CLASSIFICATION EXAMPLE PROCEDURE: AUSTRALIAN CONTEXT

In Australia, the information assets security classification procedure is guided by the Australian Government's PSPF. The PSPF provides a set of principles, protocols, and guidelines designed to help government agencies protect their people, information, and assets [46]. The PSPF, managed by the Australian Cyber Security Centre (ACSC), aims to provide a coherent approach to protective security across the government, enhancing the overall security posture of these agencies [46].

---

[1]The term ''Inventory'' will be consistently used throughout this study to maintain uniformity in the development process.

The PSPF framework is divided into four main areas, they are, Governance, which outlines security management; Personnel Security, addressing personnel security measures; Physical Security, providing guidance on securing physical assets; and Information Security, outlining a classification system for information assets based on their sensitivity and value [46], [47]].

In regard to information security, it also provides guidelines for secure information handling, access control, storage, and disposal. This ensures that information is protected from unauthorised access, disclosure, or misuse. The PSPF outlines a classification system for information assets that is based on the potential harm that could result from unauthorised access, disclosure, or misuse. There are four levels of security classification [47]:

1) Unclassified: Information that is not sensitive and can be accessed by the public without causing any harm.
2) Protected: Information that is sensitive and requires protection against unauthorised access, disclosure, or misuse. Unauthorised access to protected information could cause limited harm to national security, public safety, or the functioning of government.
3) Secret: Information that is highly sensitive and requires a higher level of protection. Unauthorised access to secret information could cause serious harm to national security, public safety, or the functioning of government.
4) Top Secret: Information that is extremely sensitive and requires the highest level of protection. Unauthorised access to top-secret information could cause exceptionally grave harm to national security, public safety, or the functioning of government.

Each level of classification has corresponding security controls and measures that must be applied to protect information. Agencies are responsible for implementing these controls in accordance with the PSPF guidelines and their own risk assessments. Enterprises also follow this framework in Australia irrespective of their relationship with the government.

In addition to the Australian Government's Protective Security Policy Framework (PSPF), other security classifications and standards are practised in the Australian industry. These often depend on the specific sector or type of data being handled. For instance, businesses in the financial sector adhere to the Australian Prudential Regulation Authority (APRA) guidelines [48]. The APRA's Prudential Standard CPS 234 [49] provides guidance on information security, covering aspects such as governance, capability, testing, incident management, and audit.

Companies that deal with cardholder data follow the Payment Card Industry Data Security Standard (PCI DSS) [50]. PCI DSS is an international standard applicable to all organizations that handle card payments, including credit, debit, and cash cards.

In the health sector, organizations comply with the Privacy Act 1988 [51], which includes the Australian Privacy Principles. These guidelines set standards for the handling of health and other personal information, including its collection, use, quality, security, and disclosure.

Businesses that operate online and handle personal information of users also adhere to the guidelines provided by the Office of the Australian Information Commissioner (OAIC) for securing personal information [52]. These include taking reasonable steps to protect personal information from misuse, interference, loss, unauthorized access, modification, or disclosure.

In essence, beyond the PSPF, Australian businesses have a range of other sector-specific and data-specific guidelines and regulations to follow when it comes to classifying and securing their information assets. The exact mix of classifications and standards to follow would depend on the nature of the business, the types of data it handles, and the regulatory context it operates in.

In the context of data levelling following the classification, in fact, enterprises follow different standards, but governments and enterprises that work with the government follow the four-level security classification discussed above. Enterprises often use a *High*, *Medium*, *Low* (or similar) classification system to simplify their data classification and risk management [53].

#### b: GLOBAL CONTEXT
Overall, there is no single global standard that explicitly sets levels for data security classification applicable across all industries and geographies. The implementation of a data classification system is usually based on the specific needs and regulatory requirements of the organization or industry.

Regulatory authorities may impose classification schemes on organisations. They can work with different types of information that have been categorised by external organisations. These include Personally Identifiable Information (PII), or data that can be linked to a specific individual; Payment Card Information (PCI), which includes credit and debit card numbers that must be handled in accordance with a payment card industry data security standard; and Protected Health Information (PHI), which includes medical records protected by HIPAA [44]. A company needs to comply with relevant industry-specific and region-specific regulatory requirements such as SOX, HIPAA, PCI DSS, and GDPR by classifying its data [54].

In summary, as cyber risks continue to grow and evolve, it is crucial for organisations of all sizes to be aware of their data's location and to classify and protect it based on its importance to their business. Numerous standards and frameworks are available for data classification, which organisations can choose from depending on their industry and specific needs, as well as regional variations.

## V. BACKGROUND ON CRYPTOGRAPHIC INVENTORY

A Cryptographic Inventory is an extensive record that encompasses all of an organisation's cryptographic assets. Such an inventory serves as a detailed map outlining the use of cryptography across various services, highlighting information on algorithms, modes of use, keys, key storage, certificates, protocol versions, library versions, and non-technical elements such as data classification, business purpose, and responsible parties [10], [55], [56].

The primary purpose of a cryptographic inventory is to provide a comprehensive understanding of an organisation's cryptographic landscape, shedding light on the strengths and weaknesses of cryptographic systems and also aiding in preparation for a potential migration to PQC. It enables organisations to grasp their cryptographic dependencies, recognise potential risks, and prioritise their cryptographic needs.

In addition, a cryptographic inventory helps organisations meet regulatory requirements by ensuring that they possess the appropriate cryptographic assets to safeguard sensitive data and comply with legal and industry standards. As a crucial cybersecurity resource, it allows organisations to implement a secure cryptographic policy in their IT infrastructure, respond promptly to security concerns, and carry out strategic transformations, such as migrating cryptographic services to the cloud or adopting PQC, with increased efficiency [56].

### A. THE IMPORTANCE OF MAINTAINING A CRYPTOGRAPHIC INVENTORY

Maintaining a cryptographic inventory is essential for a variety of reasons.

*First,* it provides a clear view of an organisation's current cryptographic landscape, which is crucial for identifying the systems and components that need to be migrated to PQC algorithms.

*Second,* PQC algorithms are still in their early stages, and the algorithms approved by the current NIST process may become vulnerable in the future, necessitating rapid changes.

*Third,* advances in cryptanalysis could compromise any algorithm, regardless of the progress in quantum computing. This has been seen with many cryptographic algorithms in the past, such as MD5 and SHA-1 [57], and more recently with an isogeny-based PQC candidate, SIKE, which has been shown to be vulnerable to a key recovery attack [32].

*Finally,* there are immediate business benefits to cryptographic inventory, such as improved crypto management, migrating to cloud infrastructure, a more robust response to incidents, meeting regulatory requirements, passing security audits, and the ability to adopt new cryptographic technology, making it a far superior strategy than solely focusing on PQC migration.

### B. CHALLENGES OF MAKING A CRYPTOGRAPHY INVENTORY

The challenges of building a cryptographic inventory for an organisation's post-quantum migration can be many-fold.

Such challenges may vary from organisation to organisation because of differences in their environment and requirements. Seven fundamental challenges that vendors, industries and standard authorities discuss are listed below [56], [58], [59], [60].

#### 1) COMPLEXITY

Organisations often have complex IT infrastructures with numerous applications, systems, and cryptographic assets. Identifying, documenting, and tracking all these assets at the same time can be difficult; such organisations may need an inventory of thousands of applications and tens or hundreds or thousands of endpoints. Therefore, scalable tools and processes that can assist with the process across the organisation are necessary. The data model used to generate and store the results is also critical. A hierarchical model can be preferable for larger organisations over a flat spreadsheet that quickly becomes unmanageable.

#### 2) HETEROGENEOUS ENVIRONMENTS AND LACK OF VISIBILITY

Organisations typically have a mix of legacy systems, modern applications, and third-party services, which can create compatibility and interoperability challenges to maintaining a comprehensive cryptographic inventory. Again, gaining full visibility into all cryptographic assets is difficult, especially when they are distributed across different departments, teams, or geographic locations, leading to incomplete or inaccurate inventories.

#### 3) CRYPTOGRAPHIC DEPENDENCIES

Understanding and developing the relationships and dependencies between cryptographic components can be challenging, requiring in-depth knowledge of the organisation's systems, data flows, and cryptographic implementation details.

#### 4) ACCESS

Inventory development necessitates collaboration from multiple organisational stakeholders, whether using cryptographic inventory tools or manual methods such as code review. Access to code, test environments, filesystems, containers in the deployment pipeline, cloud VMs and network endpoints may be needed - which can be challenging.

#### 5) HUMAN ERROR

Human errors in any manual processes involved in creating and updating a cryptographic inventory can undermine its effectiveness and impact the organisation's security posture.

#### 6) PRECISION AND THE EVOLVING CRYPTOGRAPHIC LANDSCAPE

The field of cryptography is evolving, with new algorithms, protocols, and standards emerging, making it challenging to keep the inventory updated and aligned with the latest cryptographic developments. As such, any errors (i.e. false positives or false negatives) can significantly reduce the value and effectiveness of an inventory, leading to wasted time on unnecessary remediation work, undetected crypto risks or non-compliance, and misplaced priorities for migration.

### 7) RESOURCE CONSTRAINTS AND PRIORITISATION
Due to resource limitations, cryptographic inventories may not uniformly cover all IT assets. For example, external-facing applications handling customer data almost always require more detailed analysis. In contrast, internal applications dealing with nonsensitive data may only need a basic scan where some risk of imprecision is acceptable. Since not everything can be accomplished simultaneously, it is essential to identify an appropriate roadmap with a short time-to-value to ensure buy-in and realistic completion milestones by prioritisation, which can be challenging.

### 8) CONTINUOUS UPDATES
Cryptographic systems and assets must be regularly updated to address evolving security threats and vulnerabilities. Keeping an inventory up-to-date requires ongoing monitoring, assessment, and updating efforts, ensuring that the organisation remains prepared for new challenges in the cryptographic landscape.

In general, developing a cryptographic inventory involves overcoming challenges such as complexity, scale, dependencies, heterogeneous environments, visibility, continuous updates, human error, resource constraints, and an evolving cryptographic landscape. Addressing these challenges requires strong leadership, collaboration, specialised knowledge, and a commitment to ongoing improvement.

### C. DEVELOPING A CRYPTOGRAPHIC INVENTORY
A cryptographic inventory should ideally encompass all cryptographic objects utilised across an organisation's applications, infrastructure, and networks, including those relevant to static and transit data [59]. This should include an inventory of algorithms and keys as they are employed in applications and infrastructure.

Generally, the specification of the objectives of the inventory will determine its scope [59], [60]. For instance, if the goal of the inventory is to prepare for post-quantum migration, it must differentiate between vulnerable and non-vulnerable cryptographic algorithms to facilitate planning and decision-making during migration. On the other hand, if the aim is to enforce a cryptographic policy, the inventory should be as comprehensive as the policy itself. For example, if the policy limits the use of an algorithm to a specific protocol, the inventory must include protocols. Similarly, if the policy mandates the use of certain cryptographic libraries, the inventory should include the libraries and their corresponding versions [59].

In all cases, however, it is crucial to gain a comprehensive understanding and full visibility of the application and management of cryptography within an organisation, encompassing the locations of keys and certificates as well as their maintenance across various application, computing, service, and network contexts. This can be achieved through a cryptographic inventory, as the utilisation of encryption is subject to constant change [59], [61].

Compiling a cryptographic object inventory within an organisation involves a systematic approach to identify, classify, and document all cryptographic objects used across applications, infrastructure, and networks. Such a compilation procedure can be manual, automated, or hybrid.

### 1) MANUAL METHOD OF INVENTORY COMPILATION
Manual cryptography inventory compilation involves using spreadsheets or other documentation methods to manually track and record cryptographic objects used across an organisation's applications, infrastructure, and networks. This process typically includes the following steps:

- Identify cryptographic objects: Manually search for and identify cryptographic objects within the organisation, including algorithms, keys, certificates, and libraries used in applications, network devices, operating systems, and other infrastructure components.
- Document cryptographic objects: Record the identified cryptographic objects in a spreadsheet or other document format, including relevant details such as object type, usage, location, dependencies, and responsible parties.
- Update and maintain the inventory: Regularly review and update the inventory to account for changes in the organisation's cryptographic landscape, such as the introduction of new applications or infrastructure components, updates to existing systems, and the retirement of old cryptographic objects.

Although this manual process can be suitable for small-scale applications or isolated environments, it is generally not feasible for modern enterprise teams due to its time-consuming nature, high error rate, and potential for inaccuracies. Instead, organisations should consider using automated technology to discover, track, and manage their cryptography inventory more effectively and accurately. Automated solutions can help identify key and certificate usage, streamline the inventory process, and detect anomalies, such as rogue keys, with greater precision and efficiency [60].

### 2) AUTOMATED METHOD FOR INVENTORY COMPILATION
Cryptographic inventory tools that are automated have been specifically developed to find, recognise, and report configured keys and certificates. These tools help administrators monitor the use of cryptography, while also helping to reduce the spread of keys, verify access restrictions, and guarantee prompt rotation of SSH keys [60].

Specialised software tools can automatically scan an organisation's network, servers, and applications to identify cryptographic objects such as algorithms, keys, protocols, and libraries. Examples of such tools include vulnerability scanners, configuration management tools, and cryptographic asset management solutions. Some of the approaches that the automated discovery tools can adopt are as follows [];

1) Application and source code analysis: Analyse application source code and software libraries for cryptographic objects using static and dynamic code analysis tools. These tools can help identify cryptographic implementations and usage patterns within the application code.
2) Network traffic analysis: Monitor and analyse network traffic using deep packet inspection tools, Intrusion Detection Systems (IDS), and Intrusion Prevention Systems (IPS) to detect cryptographic objects used within network traffic. These tools can help identify encrypted communication channels, key exchanges, and implementations of digital signatures.
3) System and configuration audits: Conduct regular system and configuration audits to identify cryptographic objects used in operating systems, network devices, and applications. This can be done using automated tools that can audit system configurations for cryptographic settings.
4) Log analysis: Analyse log files from servers, network devices, and applications to identify the usage of cryptographic objects. Security Information and Event Management (SIEM) solutions can help correlate and analyse log data from various sources to detect cryptographic events and patterns.
5) Endpoint security solutions: Use endpoint security solutions, such as Endpoint Detection and Response (EDR) tools, to monitor the usage of cryptographic objects on individual devices within the organisation.

By combining these technical methods, organisations can effectively track and compile an inventory of cryptographic objects, enabling them to better manage cryptographic risks and enforce security policies.

## VI. POST QUANTUM CRYPTOGRAPHIC (PQC) MIGRATION FRAMEWORK: CURRENT STATE OF THE ART

A PQC Migration Framework is a methodical approach to converting an organisation's cryptographic assets to quantum-resistant equivalents. The process entails identifying, categorising, and recording all of an organisation's cryptographic assets, such as cryptographic primitives, encryption keys, digital certificates, and other cryptographic components. Such a framework incorporates protocols and procedures to securely and efficiently manage the transition of cryptographic assets throughout their lifecycle, encompassing activities such as creating, distributing, relocating, and retiring cryptographic keys and digital certificates.

As of this date, we have identified a total of five frameworks and specific recommendations for PQC migration that propose a step-by-step transition procedure. These are ETSI [62], CARAF [63], IBM® Z [64], US Govt. Memorandums [65] and NNCSA [15].

In this section, we first outline the key features and propositions of each of these frameworks and recommendations. Following this, a comparative presentation is provided in Table TABLE 3 to highlight the motivations behind our proposed framework.

### A. ETSI TRANSITION STRATEGIES

The European Telecommunications Standards Institute (ETSI) has been actively studying quantum-resistant cryptography in preparation for the post-quantum era. With the advent of quantum computing, established cryptographic algorithms such as RSA and ECC face the potential risk of being compromised. To tackle this, ETSI set up the Quantum-Safe Cryptography (QSC) Industry Specification Group (ISG) in 2015, which explores quantum-safe cryptographic solutions and proposes migration strategies [66].

ETSI has produced various reports and technical specifications that address quantum-safe cryptography and transition strategies. Two notable publications, ETSI TR 103 616 [67] and TR 103 619 [62], detail quantum-safe cryptographic schemes and mitigation strategies, providing recommendations for quantum-safe applications.

ETSI TR 103 616 evaluates the security, performance, and other attributes of NIST-announced PQC primitives, analysing their potential vulnerabilities and possible countermeasures against quantum threats. ETSI TR 103 619 discusses the transition to a Fully Quantum-Safe Cryptographic State (FQSCS), focusing on mitigating quantum computing threats to cryptographic elements.

The document offers a three-stage framework-comprised of inventory compilation, migration planning, and execution-to guide organisations toward a quantum-safe environment. ETSI proposes a comprehensive questionnaire that covers risk, data, cryptographic, infrastructure, and supplier assessments to aid in the migration process.

### B. CARAF

Chujiao Mia [54] proposed a 5D framework (named CARAF) for organisations to assess and mitigate the risks associated with inadequate crypto-agility. Their research focuses on how quantum computing may threaten cryptography, particularly in the Internet of Things (IoT) domain.

Phase 1 of the framework involves explicit threat identification, which sets it apart from other risk frameworks that we identified. Assets that are not affected by the identified threat are excluded from consideration, allowing the risk assessment process to be optimised.

In Phase 2, several factors are used to compile an inventory of assets vulnerable to security risks. These include asset scope, sensitivity, cryptographic solutions, secrets management, implementation, ownership, location, and lifecycle management. This process enables prioritisation for risk mitigation and the development of a roadmap for implementing risk mitigation strategies.

In Phase 3, the inventory is prioritised based on risk exposure, and risk estimation is performed. They propose a novel risk estimation formula, "`Risk = Timeline × Cost`," in the context of quantum computing threats to cryptography, which provides a more meaningful risk measurement.

Phase 4 involves implementing risk mitigation strategies. Organisations have three primary options: allocate resources for asset protection, accept the risk and keep the status quo, or discontinue use of affected assets.

Phase 5 concludes with the development of an organisational roadmap. This roadmap addresses crypto agility issues based on the risk mitigation strategy chosen. It relies on establishing foundational structures such as a clear crypto policy, a well-defined RACI matrix, and integration with related organisational processes and technology for greater agility. The crypto agility remediation roadmap calls for crypto policy updates to phase out deprecated algorithms and implement replacements while leveraging related processes to enforce requirements.

### C. IBM®
IBM® [64] has launched a major initiative to develop and improve cryptographic security for its IBM®Z mainframe computers. IBM®began the initiative by developing a method to examine the existing cryptographic usage on its platforms and devise strategies to fill any gaps. They collaborated with IBM Zurich Research to create a comprehensive questionnaire that focused on nine areas of cryptography and cybersecurity. Following an evaluation of the responses, they proposed preliminary plans for enhanced cryptographic security.

The IBM®Z team follows an iterative review process and keeps the cryptographic inventory up to date. They then prioritised changes and created a multiphase roadmap, focusing first on critical areas. The roadmap was influenced by algorithm recommendations, standards, guidelines, and the availability of quantum-safe hardware.

IBM®also created data and cryptographic inventories to keep track of the company's critical data assets, and it used cryptographic objects. The goal is to provide comprehensive information on data protection requirements and implementation timelines [68].

### D. USA GOVT. MEMORANDUM
The Executive Office of the President's memorandum from the Office of Management and Budget, on November 18, 2022, provides clear and actionable guidance for federal agencies in the United States, requiring them to follow National Security Memorandum 10 (NSM-10) [65].

This decree highlights that preparing for PQC is a critical component as federal agencies transition to a more secure zero-trust architecture. The emphasis is on adopting a robust approach to accomplish this and it requires agencies to conduct a comprehensive and prioritised inventory of their cryptographic systems. High-Value Assets (HVAs) and systems with high impact are given special attention because they pose a significant potential risk.

The inventory should be comprehensive and detailed, including Federal Information Security Modernisation Act (FISMA) system identifiers, FIPS system categorisation, cryptographic algorithm specifics, data lifecycle characteristics, and relevant vendor information. This level of transparency and detail will allow for a better understanding of the systems in use and more effective risk management.

By outlining this process, the memorandum provides a clear path forward and a strategic guide for agencies to safeguard critical assets during this transition period.

### E. NNCSA HANDBOOK
The Netherlands National Communications Security Agency (NNCSA) has developed a Handbook where they have outlined a framework to assist and guide organisations managing data with a long-term confidentiality requirement or those using long-lived systems to start implementing mitigating measures amid the quantum threat. They have speculated and claimed that this transition is likely to be time and resource intensive and could take more than five years.

In their framework, they have adopted a 3 step process, wherein in the first phase, organisations are advised to initiate this process by performing a PQC Diagnosis to evaluate their data handling procedures and risk exposure. This would help formulate an appropriate response strategy. The next phase involves creating an inventory of the cryptographic assets in use, the data they protect, and the individuals responsible for managing them.

Finally, as the third and final phase, the PQC migration plan was carried out. In this execution phase, they emphasise that organisations must be careful not to introduce new risks.

The Handbook underscores the importance of maintaining cryptographic agility to quickly adapt to any potential vulnerabilities or improvements in cryptographic protocols that emerge during or after migration.

## VII. PROPOSED FRAMEWORK FOR PQC MIGRATION
This section describes our proposed PQC Migration Framework, which is intended to provide organisations with a structured approach for identifying, managing, and transitioning their cryptographic assets and dependencies in preparation for a successful transition to quantum-resistant cryptography.

It begins by discussing the scope of the proposed framework, its guiding principle, and the specifics of its design and development process. Adoption of this framework would allow organisations to accurately evaluate cryptographic dependencies, plan migration to advanced cryptographic schemes, and ensure a secure and controlled transition to PQC.

### A. SCOPE
This framework focuses on determining whether the quantum security level of the cryptographic algorithms used is sufficient for the security requirements of individual data assets. It does this by evaluating the dependencies of data assets on cryptographic objects. That is to say; it analyses

**TABLE 3.** Comparison of the existing and proposed frameworks for Post-quantum Cryptography (PQC) migration based on their six most significant characteristics & features. In this table, each row represents a distinct framework, while columns detail attributes like crypto inventory methods, source consideration for cryptographic objects, and risk assessments. Cell values, ranging from "Yes", "No", to specific details, signify whether a framework incorporates a particular attribute, with "No" indicating the absence or non-specification of that attribute. Some cells provide further details about the method employed for a given attribute. For instance, "questionnaire method" implies that the cryptographic compilation is questionnaire-based, while "tool-based method" indicates the use of specific tools and methods other than questionnaires for compiling cryptographic objects.

| Framework/ Recommendation [Year] | Type of Crypto Inventory Compilation Method | Consider Source of Cryptographic Objects | Cryptographic Assessment (Crypto Object Categorisation) | Data Inventory & Security Classification | Quantum Risk Assessment | Dependency Assessment Process |
|---|---|---|---|---|---|---|
| ETSI [2020] | Questionnaire | Yes (Supplier & Infrastructure) | Yes (Separate process [69]) | Yes | Yes | None |
| CARAF [2021] | None | No | No | No | Yes | No |
| IBM Z [2022] | Tool-based + Questionnaire | No | Yes (Questionnaire) | Yes | Yes | Partial (Proprietary tool) |
| USA Govt. [2022] | None | No | Yes | Yes | Yes | No |
| NNCSA [2023] | None | No | Yes | Yes | Yes | No |
| **Our proposal** | **Tool-based** | **Yes** | **Yes** | **Yes** | **Yes** | **Yes** |

* No → Not Specified

which cryptographic objects - such as encryption keys, digital signatures, and cryptographic protocols - contribute to securing each data asset within the context of data storage, transmission, and processing. Using this information, the framework provides a methodology for discovering problematic uses of quantum unsafe cryptography. This information can further be used in risk analysis and migration planning.

It is important to note that this framework is specifically centred on the dependencies of cryptographic objects and does not cover certain aspects of information security management. For example, the framework does not encompass a detailed risk assessment, business plans, identifying threats, or determining the most appropriate type of cryptographic primitives (e.g., digital signature vs. encryption). As a result, this framework should be utilised in conjunction with other relevant security frameworks and methodologies, such as NIST CSF [70], ETSI [62] or CARAF [63] etc., that address these out-of-scope areas to ensure a comprehensive approach to data security for post-quantum migration.

### B. DESIGN PRINCIPLES

The design principles for the framework provide precision, flexibility, and security. This approach aims to facilitate a smooth transition from traditional cryptographic mechanisms to post-quantum alternatives while adapting to the evolving quantum computing threat landscape.

- **Precision** in identifying dependencies of cryptographic objects with the assets is crucial to understand the specific relationships between cryptographic capabilities and their influence on data security, enabling organisations to make informed decisions on selecting and implementing post-quantum cryptographic solutions that are required and best address their needs.
- **Security** remains the primary part, with the framework designed to identify, evaluate, and implement post-quantum cryptographic solutions that provide robust protection against quantum attacks. This process includes assessing the current cryptographic inventory and vulnerabilities to develop crypto agility and offering recommendations for selecting and implementing suitable post-quantum alternatives aligned with specific data security requirements.
- **Flexibility** ensures the framework's adaptability to new developments in quantum-resistant cryptographic algorithms and standards and its compatibility with other security frameworks and methodologies, making it suitable for organisations of all sizes and industries. This allows for seamless integration with existing security infrastructures and the ability to address out-of-scope areas such as risk assessment, threat identification, and business planning.

### C. FRAMEWORK DESIGN

In order to assist organisations in identifying, evaluating, and migrating their cryptographic systems in preparation for post-quantum computing, our Cryptographic Inventory Framework provides a structured methodology. The structure is comprised of the following six primary processes as shown in FIGURE 5 and introduced below:

### 1) CRYPTO INVENTORY PROCESS

The first step in the Inventory Process is to compile an exhaustive inventory of the cryptographic systems, algorithms, and protocols currently used throughout the organisation. This includes both the software and hardware components, as well as the communication channels and services provided by third parties. After a thorough understanding of the organisation's cryptographic landscape has been attained, potential vulnerabilities can be located.

A variety of methods that can be employed in the compilation of the inventory are presented in detail in Section V-C. Manual Collection necessitates a systematic but labour-intensive inventory of cryptographic resources. In contrast, Automated Collection, Network Scanning, and cloud-based systems leverage software, platforms, and real-time data to detect, catalogue, and monitor cryptographic assets across networks and multi-cloud environments. Public Key Infrastructure managers extract information about keys and certificates. These combined methods facilitate an organisation's readiness assessment and strategic planning for Post Quantum migration.

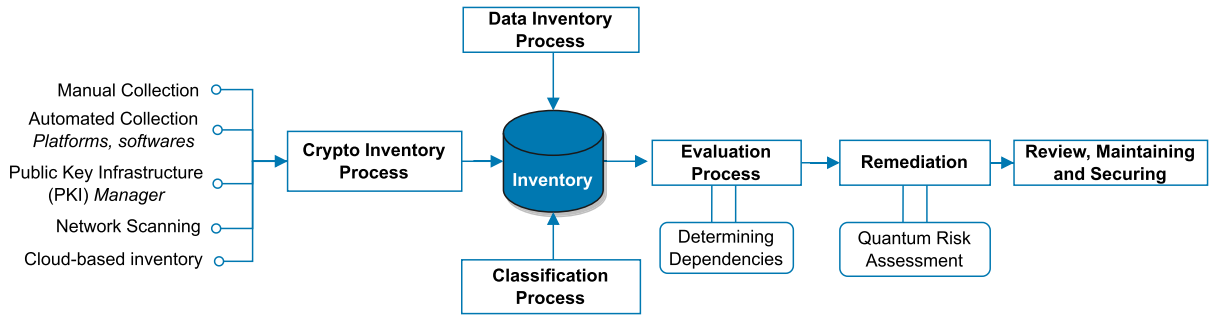

**FIGURE 5.** Proposed Framework Process for building cryptographic inventory for post-quantum migration.

#### 2) DATA INVENTORY PROCESS:

At this stage, organisations are responsible for cataloguing sensitive data that is protected by cryptographic mechanisms.

The initial step in this process is to determine the different types of data and their locations before learning how the data move throughout the organisation. In general, organisations nowadays are at a critical juncture in their journey through the complex tapestry of information management, where they must systematically catalogue sensitive data protected by cryptographic measures. To begin this complex endeavour, the first step is to identify not only the various data types but also their precise repositories. Certain confidential financial records, for example, may be stored on secure on-premises servers, whereas encrypted employee communications may be stored in cloud-based solutions such as Microsoft Azure or Google Cloud Storage.

Understanding data mobility is critical; this includes mapping out pathways such as data relayed from regional offices to headquarters or from secured databases to frontline applications. As the migration phase approaches, the focus shifts to determining which of these data assets is critical. The goal here is to fortify these prioritised assets by enclosing them behind the organization's most impenetrable cryptographic defences. In Section IV-A, a general discussion on the recording of enterprise data (information) assets was previously presented.

#### 3) CLASSIFICATION PROCESS

Following the completion of the inventory, the next step in the classification process involves categorising the organisation's data assets according to the risk levels they present and the potential damage that could result from a breach in security.

For example, financial statements or proprietary algorithms, due to their essential nature, may be classified as 'high risk' owing to the significant consequences that could arise from a potential breach, such as substantial financial losses or competitive disadvantages. On the other hand, it is possible to classify generic company newsletters as having a relatively low level of risk. In addition to the inherent sensitivity of the data, compliance also assumes significance. Data that falls under stringent regulations such as the General Data Protection Regulation (GDPR) or the Health Insurance Portability and Accountability Act (HIPAA) necessitates special consideration due to the possibility of legal consequences and substantial penalties for different parts of the world.

The evaluation of these factors serves the dual purpose of illuminating the vulnerability landscape and facilitating the optimisation of the migration strategy. Through the strategic utilisation of this classification system, organisations are able to intelligently allocate resources and time, thereby prioritizing the migration of the most crucial and high-risk systems and data assets.

In previous Section, Section IV-B, more general discussion on the classification and the specific context of Australia is presented as an example.

#### 4) EVALUATION PROCESS:

The evaluation process begins with this phase, in which organisations determine the degree to which each cryptographic system and data asset depend on traditional cryptographic algorithms, which can be susceptible to quantum attacks. This evaluation process helps organisations plan for replacements based on their specific use cases and security requirements.

This process primarily focuses on determining dependencies, a comprehensive assessment of the interdependencies and relationships between enterprise IT assets, data assets, and their underlying cryptography. This evaluation considers various facets of dependencies to ensure a more robust security stance against potential quantum attacks by critically examining how data assets rely on cryptographic elements. A detailed presentation on how to determine cryptographic dependencies is presented in Section VIII. The evaluation process thus facilitates organisations in selecting suitable replacement algorithms tailored to their specific use cases and security requisites. It allows the identification of systems that necessitate immediate migration to post-quantum cryptographic algorithms and highlights the urgency of transitioning these systems to secure against quantum threats.

Furthermore, within this scope, enterprises can conduct benchmarking to compare the performance of potential PQC algorithms to current cryptographic solutions. This includes processing speed, data overhead, compatibility, and other metrics that may have an effect on system performance or user experience.

Also, before full-scale implementation, enterprises may consider pilot tests of PQC on less critical systems to evaluate the system and identify potential pitfalls in order to aid the streamlining the migration process. Within this scope, enterprises can expand their investigation into identifying the vulnerabilities of emerging PQC schemes recommended by NIST for their own use.

Finally, the evaluation can be expanded to include enterprise-wide Policy & Protocol Revision in order to update organisational policies, protocols, and guidelines to reflect the shift to PQC. Modifications to data storage and transfer protocols, password and key management, and incident response strategies may be required.

### 5) REMEDIATION PROCESS
The actual transition of vulnerable cryptographic systems to PQC is the focus of this stage of the remediation process. The Remediation Process involves developing a roadmap for PQC migration and implementing changes to the cryptographic assets identified in the Classification and Evaluation Processes. It includes bringing cryptographic libraries, hardware, and software components up to date, implementing new cryptographic algorithms and protocols, testing and validating the security and functionality of the updated systems, and training employees on the changes and new security best practises.

As part of this process, Quantum Risk Assessment is integrated into the step. Each individual organization will conduct their own risk assessment to determine the gravity of quantum risks and to prioritize the migration according to their specific needs. This will be accomplished in alignment with the timeline recommended by NIST and by adopting well-established risk assessment techniques delineated in other frameworks.

### 6) REVIEW, MAINTENANCE, AND SECURITY
After a successful migration, it is paramount for a company to transition into a phase of continuous vigilance. The cryptographic landscape is dynamic, and what's secure today might not be tomorrow. Therefore, routinely reviewing and updating the framework of the cryptographic inventory is essential to ensure its accuracy and relevance. For instance, the quantum threat to Post-Quantum Cryptography (PQC) serves as a prime example of the evolving challenges in the cryptographic domain. As quantum computing advances, new algorithms emerge that can potentially compromise even the PQC schemes. Regularly reviewing such threats ensures that the company's cryptographic measures remain robust against the latest quantum challenges.

Beyond this, it is crucial to conduct regular audits, vulnerability assessments, and risk assessments. These activities help in identifying new vulnerabilities or threats that might have emerged post-migration. A proactive approach to security doesn't stop at identification; it extends to devising preventive maintenance strategies. This ensures the continued safety and integrity of the company's cryptographic systems. Staying updated with the latest advances in PQC and other cryptographic fields is also vital. As the field evolves, businesses should be ready to proactively update their software and systems, ensuring they remain at the forefront of cryptographic security.

Most of the above processes will follow established processes and may already be implemented in organisations. Our main contribution is in the evaluation process, discussed in the next section.

## VIII. DETERMINING DEPENDENCIES
This section starts delving into the fundamental rules for identifying security dependencies within organisations, an important contribution of our presentation. Using graph theoretic techniques, we uncover potential security challenges. While we have touched upon the rules, it is imperative to elucidate how these rules are applied in real-world scenarios. Three case studies are presented to provide a comprehensive understanding. These studies are intended to depict real-world scenarios, demonstrating how cryptographic dependencies manifest in various environments, ranging from the lifecycle of a classified document to the complexities of a hybrid IT infrastructure.

The focus of **Case Study 1** is a classified 'secret' document. We conduct a thorough examination of its cryptographic dependencies, spanning three critical states: at rest, in motion, and in process. This emphasises the critical importance of safeguarding sensitive business policy statements.

**Case Study 2** delves into an enterprise's operations within a hybrid IT infrastructure. In this section, we navigate the cryptographic nuances associated with storing sensitive customer data on-premise and via third-party cloud backups.

Finally, in **Case Study 3**, we provide a brief example set within the cloud ecosystem. This study highlights various types of security dependencies. We highlight the use of various enterprise artefacts and inventories, such as security classifications, data sets, cloud configurations, and crypto inventories, which are all derived from a variety of inventory types.

### A. SECURITY DEPENDENCY RULES OF THUMB
Here, we outline some basic rules of thumb for finding security dependencies within an organisation.

### 1) SECURITY LEVELS

Security levels refer to properties of cryptographic primitives and protocols that could be required for specific data classifications. For example, it might be required that primitives are NIST-approved or quantum-safe. We can also specify a particular quantified security level, such as 128-bit quantum security.

Security levels are vertices in the graph, just like assets.

- A security level depends on the data classifications that require it (e.g. if top secret requires 256-bit security then 256-bit security depends on top secret)
- Cryptographic algorithms depend on the security levels that they have as properties (eg. AES-128 depends on 128-bit security.)

The above rules allow us to use standard graph theoretic techniques to find potential security problems, for example, by finding a path from a security level to a lower or incompatible security level. The rules are perhaps a bit counter-intuitive at first, but it is helpful to think of a security level vertex as representing a property that cannot be contradicted by any descendent vertex. For example, drawing an edge from 256-bit security to top secret indicates that top secret (and all its descendants) must not contradict 256-bit security, for example by using AES-128 which implements 128-bit security.

### 2) DATA CLASSIFICATIONS

- A classification depends on any data that has it as a classification (e.g. if D is classified as top secret, then top secret depends on D).

### 3) DATA

The dependencies for data are straightforward and follow the standard security practise of considering whether data are at rest, in transit, or in use.

- Data depends on any asset that it is stored on.
- Data depends on any asset (eg. communication channel) that it is communicated over.
- Data depends on any asset (eg. process) that makes use it.

### 4) CRYPTOGRAPHIC KEYS

Cryptographic keys are important in that they form a bridge between processes and cryptographic primitives. In public key cryptography there are also relationships between public and private keys. Finally, they are also a kind of data. Therefore, keys can be complex dependencies.

- Secret (symmetric) keys depend on any asset that it is stored on, communicated over, or used with (eg process that might leak the key through a side channel).
- Private keys are treated the same as secret keys.
- Public keys depend on their matched private key (the public key provides no security if the private key is leaked) and any asset stored on (a corrupted public key can compromise authentication).

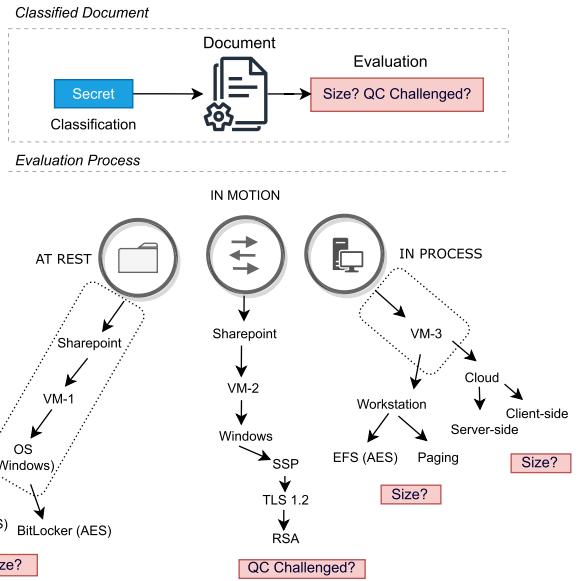

**FIGURE 6.** The Evaluation Process of a classified document.

- All keys depend on the cryptographic primitive that they are used with. If the primitive is broken, the key provides no security.
- All keys depend on the process that created them.
- Certificates depend on the algorithm for their signature, on the public key, and on the certificate of the CA, unless self-signed.
- CA certificates additionally depend on the asset on which they are stored.

### 5) CRYPTOGRAPHIC PROTOCOLS AND PRIMITIVES

Cryptographic protocols and primitives have an additional layer of complexity in that they can have multiple key sizes or configurations. For example, TLS can be configured to allow or disallow a wide variety of cipher suites. Accordingly, we need some additional information when determining the dependencies of these objects. Ideally, these dependencies, based on possible configurations, will be accessible through a publicly available database as they do not vary from organisation to organisation.

- Any asset (e.g. cryptographic primitive) that provides security depends on its security level (eg RSA-512 depends on Low). This rule allows us to find directed paths from high security levels to lower or contradicting security levels through insecure cryptography.
- A protocol that uses a primitive depends on the primitive.
- If a protocol can have multiple configurations then each protocol/primitive pair will depend on the primitives available in that configuration.

### 6) MACHINES, VIRTUAL MACHINES, CONTAINERS, ETC. (PROCESSORS)

There are multiple types of assets that process data, from physical computers up to serverless functions. We can treat them similarly, although smaller assets like containers and serverless functions will typically interact with fewer data assets and hence allow for a more fine grained analysis.

- A processor depends on any symmetric keys that it stores. If the key is compromised, then any cryptographic processing performed does not provide security.
- A processor depends on any public keys or CA certificates that it stores. Compromised public keys or CA certificates can lead to compromised authentication.
- A processor can be treated as a proxy for a process if no inventory information exists on the processes that it runs (e.g. container running a web server where the web server does not appear in an inventory but the container does).

### 7) PROCESSES

Processes include software, libraries, workflows combining other processes, and business processes. This category can also include other entities like processors that are proxies for processes or perform similar functions.

- A process depend on cryptographic protocols or primitives that they use.
- A process depends on the processor that it runs on.
- A process depends on any software, libraries, subprocesses, etc. that it uses. This may be configuration/ workflow dependant, (e.g. making use of an encryption option when saving a document).
- A process depends on any cryptographic key (including public keys) that it uses.

### 8) COMMUNICATION

In many cases there may not be direct information on communication channels, but they are useful in establishing dependencies between entities that communicate data to other entities.

- A communication channel depends on any cryptographic protocols, primitives, or keys that underlie it.
- A communication channel has a two-way dependency, with each entity communicating over the channel.

Using a variety of scenarios, the following sections evaluate various dependencies in relation to cryptography in enterprise settings.

### B. CASE STUDY 1:
### DEPENDENCY EVALUATION OF A CLASSIFIED DOCUMENT

This case study details the analysis of cryptographic dependencies associated with a classified 'secret' Microsoft Word document containing sensitive business policy statements in a typical windows-based business environment. In such cases, the cryptographic dependencies should examine in three states of the document- at rest, in motion, and in process as shown in the FIGURE 6.

We consider that the document under evaluation is stored in SharePoint. The environment is primarily windows-based, including standalone workstations and shared spaces on SharePoint. Encryption solutions like Encrypted File System (EFS) and BitLocker are employed for data protection. Note that the process should be the same for other operating system-based environments, such as Linux.

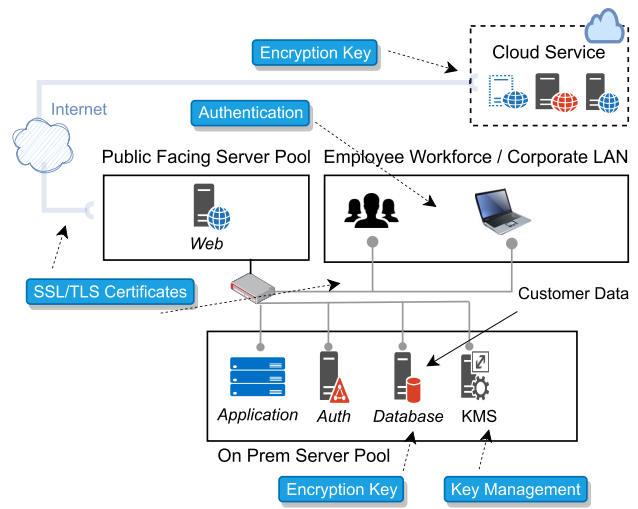

**FIGURE 7.** Identified Enterprise Assets from a part of a hypothetical enterprise network, that uses different cryptographic objects.

### 1) AT REST

The first evaluation phase, focused on the document at rest, is stored on a device using a SharePoint application within a popular Windows-based environment. We first consider the filesystem-level encryption provided by the Encrypted File System (EFS), which could be using the AES algorithm. Assessing the security of the encryption involves examining the cryptographic keys used by EFS and the security of the key management infrastructure. The key assessment should consider the strength of the keys (e.g., length, randomness), and how they are generated, stored, and protected.

When considering full-disk encryption solutions like BitLocker, it is important to identify the underlying cryptographic algorithm. BitLocker typically uses AES with a 128-bit or 256-bit key in the Cipher Block Chaining (CBC) mode, where an initialization vector is used to ensure that identical plaintext blocks are encrypted into different ciphertext blocks. Like with EFS, the cryptographic keys' strength, storage, protection, and quantum resilience are crucial.

### 2) IN MOTION

The next phase addressed the document in motion. This has to start with identifying the underlying operating system (OS) and its communication security provisions. In this scenario, Windows employs Security Support Provider (SSP), where it can use the Transport Layer Security (TLS) 1.2 protocol, potentially using RSA encryption for key exchange. Despite using RSA for key exchange, we note that RSA is vulnerable to quantum threats.

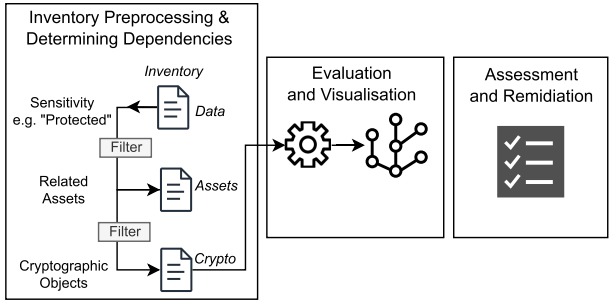

**FIGURE 8.** Evaluation Process utilising different inventories.

### 3) IN PROCESS

Finally, in the "in process" state, the document is actively in use or being processed. Depending on the environment-local or cloud-the employed cryptographic objects have to be examined. If it is a standalone Windows workstation, the cryptographic technique of EFS must be identified, like the key management scheme, encryption algorithm, and its quantum resilience.

For a cloud environment, a deeper analysis is needed. The client-side and server-side encryption techniques, both in transit and at rest, need to be verified for robustness and quantum safety. Additionally, a holistic understanding of the shared responsibility model in cloud security is necessary to ensure the proper implementation of security controls.

This multi-state evaluation allows us to identify the cryptographic dependencies of the document and assess the risk based on the CIA triad. In this way, we can ensure that the document's confidentiality, integrity, and availability are preserved while mitigating quantum and other emerging threats.

### C. CASE STUDY 2:

#### DEPENDENCY EVALUATION OF A CORPORATE DATABASE

##### a: SCENARIO DESCRIPTION

In this case study, we examine a scenario in which an enterprise operates in a hybrid IT infrastructure with sensitive customer data stored in an on-premise secure database. The enterprise also uses a third-party cloud service for data backups. The scenario can be visualised as shown in FIGURE 7.

This diagram displays a number of servers in the On-Prem Server pool, including one that hosts the customer database used by the Customer Relationship Management Software (CRM software). This CRM software is served from the application server of the company's enterprise system. Employees from the corporate LAN interact with the database through the CRM software. Customers, on the other hand, perform Create, Read, Update, and Delete (CRUD) operations through a web server located within a public-facing network. Additionally, an automatic backup system is in place, linked to a cloud service for data storage and retrieval.

Therefore, within this context, three key access processes are identified: customer access in CRM software through

**TABLE 4.** Snippet of data inventory with fields and values.

| Field | Value |
|---|---|
| Data Identifier | DA005 |
| Data Name | CustomerData |
| Asset Identifier | PA001029 |
| Asset Name | CRMDatabase1 |
| Data Category | Customer Records |
| Sensitivity | Protected |
| Storage Location | On-premises |
| Encryption in Use | AES-128 |
| Legal and Regulatory Requirement | Privacy Act 1988 (Cth) |
| Retention Period (Years) | 10 |

**TABLE 5.** Snippet of the asset inventory with considered fields and values.

| Field | Value |
|---|---|
| Asset Identifier | PA001010 |
| Asset Name | WebServer1 |
| Device Type | Server |
| Location | Building A |
| Provider | Dell |
| Model | PowerEdge R740 |
| Memory | 128 GB |
| IP Address | 192.168.1.2 |
| Serves | CRMSoftware1 |
| Database Server | - |

**TABLE 6.** Snippet of the cryptographic inventory with the fields and values.

| Field | Value |
|---|---|
| Crypto Identifier | CR001 |
| Asset Identifier | PA001010 |
| Asset Name | WebServer1 |
| Cryptographic Object | WebServer1_SSL_Cert |
| Object Type | SSL/TLS Certificate |
| Algorithm | RSA |
| Key Size | 1024.0 |
| Key Location | KMS |

Web servers where the CRM software's back-end is an on-prem database. Employees have access to the database through the corporate network through the CRM software, and a backup of the database is uploaded to third-party cloud storage.

In these three processes, the confidentiality, integrity, and authenticity of the data are safeguarded by various cryptographic techniques, establishing dependencies on different cryptographic objects and processes.

##### b: DEPENDENCIES

In this case study, the dependencies that are evaluated refer to the reliance of enterprise systems, software, or assets on cryptographic algorithms or cryptographic objects for their security and functionality. This might involve using encryption and decryption methods, digital signatures, hash functions, random number generators, and so forth. It also refers to the reliance on the cryptographic objects of one asset to other related assets that have access permission within them.

These dependencies are illustrated in FIGURE 7 and described as follows:

1) Database Encryption: The enterprise employs symmetric encryption algorithms to encrypt the customer data at rest in its databases. This encryption relies on cryptographic keys that must be securely managed and stored. The confidentiality and integrity of the customer data are contingent upon the security of these keys, key sizes, and key management systems.

2) Secure Communication: The enterprise's web servers establish secure network connections with the databases using SSL/TLS protocols, ensuring encrypted and authenticated communication. This secure communication relies on SSL/TLS certificates, which serve as cryptographic objects. The confidentiality and integrity of data in transit, depending on the security of these certificates, may depend on the cryptographic library and KMS.

3) User Authentication: Employees access the enterprise's systems through secure login processes that utilize cryptographic objects such as hashed and salted passwords, and potentially digital certificates for two-factor authentication. The security of the systems and the data they access depends on the integrity of these cryptographic objects.

4) Cloud Services: The enterprise leverages a third-party cloud service for data backups. Data is transferred to the cloud service over a secure connection, once again relying on SSL/TLS certificates. Additionally, considering the client-side encryption technique and server-side encryption technique the enterprise can encrypt the data before sending it through and also the cloud service encrypts the data at rest, creating a dependency on the encryption keys managed by the enterprise system and the cloud provider, respectively.

5) Key Management: The cryptographic objects mentioned above, including the database encryption keys, SSL/TLS certificates, hashed passwords, and cloud provider's keys, must undergo secure management. This often involves a key management system or service, which in turn relies on cryptographic objects and processes for its secure operation.

Understanding and evaluating these dependencies is crucial when considering the migration to PQC to ensure the continuity of security measures. By analysing the relationships between sensitive assets and their cryptographic dependencies, organisations can effectively plan for the adoption of PQC and mitigate potential security risks.

### c: EVALUATION
The identification and evaluation of cryptographic dependencies tied to various assets can be accomplished through the use of enterprise artifacts, such as comprehensive enterprise-wide lists of diverse assets. In this specific context, we focus on three types of inventories [2]: a) a data inventory, which contains all data along with their attributes and sensitivity classifications; b) an asset inventory, which enumerates all the software and hardware assets; and c) a cryptographic inventory, which provides a list of all the cryptographic objects utilised within enterprises.

The efficient use of these inventories allows for the identification of cryptographic objects and their associated elements within this process.

The first step in dependency determination involves identifying data assets with respect to their sensitivity from the data inventory, for example, that are marked as 'Protected'. As part of the standard risk management plan, higher sensitive data must be secured and prioritised during PQC migration. Once the asset is identified, other connected assets can be discerned from the asset inventory. This can be achieved through a process of nested iterations: for each high-sensitivity asset, corresponding assets are identified from the asset inventory.

Subsequently, for each linked asset, associated cryptographic objects from the crypto inventory are identified, culminating in a comprehensive exploration of cryptographic dependencies through evaluation and visualisation which will be ready for PQC assessment and remediation, as shown in the FIGURE 8.

In our demonstration, we prepared three datasets, namely data inventory, asset inventory, and crypto inventory, and assumed they were available from the organisation in Excel format. A snippet with typical fields with items for data, asset and crypto inventory is given in the TABLE 4, TABLE 5, and TABLE 6.

The dependencies are generated and presented using the principles of directed graph theory. In fact, by using this method, we can effectively capture the complex web of dependencies that exists within our datasets. We used NetworkX, a well-known Python library, to implement this representation. Algorithm-1 describes the underlying algorithm that powers this representation. FIGURE 9 depicts these dependencies as the general output of Algorithm-1.

As the part of the process in this scenario, the algorithm extracts dependencies from three distinct datasets. These interdependencies are then represented on a directed graph. The nodes in this graph are symbolic representations of various assets, data sets, and cryptographic objects. These are the entities that are dependent on one another. The edges, on the other hand, serve as connectors, representing the nodes' actual dependencies or relationships. In essence, an edge from one node to another indicates that the former is in some way dependent on the latter.

This visualisation not only provides a snapshot of the dependencies, but it also helps to understand the flow and hierarchy of dependencies, making it easier to identify potential issue or critical nodes of particular interest. The

---

[2]Following on the terminology discussion from Section IV, we are referring to these artifacts as inventory.

---

**Algorithm 1** Algorithm for Evaluation and Visualisation of Dependencies in Case Study 2

---

**Input:** "data_inventory", "asset_inventory", "crypto_inventory"

**Output:** Asset-Cryptography Dependency Evaluation

1 Filter 'Protected' sensitivity data from data_inventory_df to high_sensitivity_assets

2 Initialise an empty directed graph, G

3 **foreach** *row in high_sensitivity_assets* **do**

4     Extract 'Data Identifier' and 'Asset Identifier'

5     **if** *they are not Null and not equal* **then**

6         Add edge in G; Find related_assets in asset_inventory_df

7         **foreach** *row in related_assets* **do**

8             Extract 'Asset Identifier'

9             **if** *it and original 'Asset Identifier' are not Null and not equal* **then**

10                 Add edge in G; Find related_crypto_assets in crypto_inventory_df

11                 **foreach** *row in related_crypto_assets* **do**

12                     Extract 'Crypto Identifier', 'Algorithm', 'Key Size', 'Key Location'

13                     **foreach** *unique pair among extracted values* **do**

14                         **if** *not Null and not equal* **then**

15                             Add edge in G

16                       **end**

17                   **end**

18               **end**

19             **end**

20         **end**

21     **end**

22 **end**

23 visualise G

24 **End of the algorithm**

---

output visualising the dependencies in FIGURE 9 are as follows;

- The database node has outgoing edges to the encryption keys (indicating a dependency on the keys for secure storage) and to the database server (necessary for data management and providing access, security, etc.).
- The web server node depends on SSL/TLS and SSL/TLS depends on KMS, so the web server has an outgoing edge to the SSL/TLS certificates, SSL/TLS has an outgoing edge to the KMS (necessary for secure communication).
- The cloud service node has outgoing edges to its own encryption keys and to the SSL/TLS certificates (necessary for secure communication).

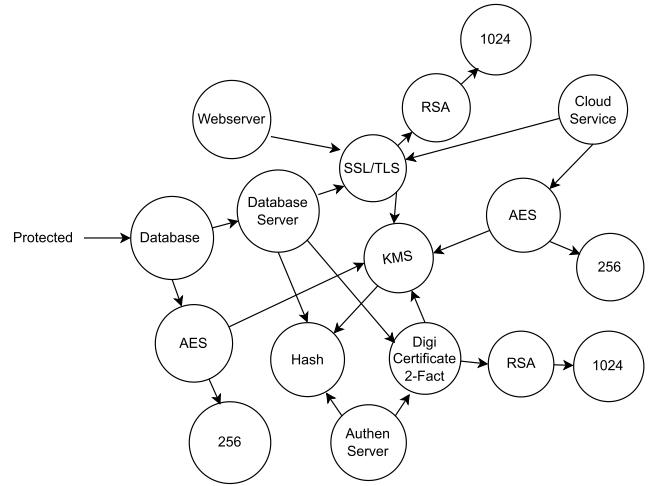

- The user authentication node has outgoing edges to the hashed passwords and digital certificates.
- The key management system node receives incoming edges from all other nodes that rely on the cryptographic objects it supervises.

Such an approach of determining and evaluating dependencies is critical to the migration process towards the PQC era. This graph facilitates the visualization of dependencies and possible system security weak points. For instance, if the SSL/TLS certificates were compromised, the security of both the database and the cloud service would be jeopardized, potentially affecting other systems as well. Similarly, a breach in the key management system could detrimentally impact the security of the entire enterprise's data.

### D. CASE STUDY 3:
### A DETAILED MINIMAL EXAMPLE

In this section, we present a minimal example in the cloud ecosystem that illustrates different types of security dependencies utilising different enterprise artefacts and inventories, such as security classification, data, cloud configuration, crypto inventory, etc., that can be gleaned from various types of inventory.

The dependencies are presented as short tables mapping the various relationships that need to be tracked. Importantly, we identify security dependencies by the *type* of information contained in each row/record. Thus, although we have only one row/record in each inventory, the same pattern of security dependencies can be applied automatically to every row/record in the inventory. So, although there is some manual process required to identify the type of dependencies present in the inventory, this needs to be done only once per inventory. The actual dependencies can be extracted automatically, and the manual work does not need to be re-done if the inventory contents (not structure) change.

*a: INVENTORIES*

Here we provide the inventories used in this example. We have provided only the columns that we used for establishing dependencies and omitted all other information that would normally be present in these types of inventories.

1) **classifications.csv**

   This file provides a mapping between organisational data classifications and security levels. This is necessary since organisations follow a wide variety of classifications and have varying requirements and risk tolerances. Meanwhile, cryptographic primitives can be rated in several ways, for example, according to the effort required to break them, or by whether they are considered quantum-safe, or approved by a standards body such as NIST. Here, there will be a dependency from the security level (NIST approved in this example) to the classification.

   ```
   Classification,Security
   High,NIST-approved
   ```

2) **data.csv**

   This is a data inventory which is a common type of inventory. Here, we have included storage location and classification in addition to the asset identifier. According to the rules of thumb, there will be dependencies for each row from the Classification to the ID and from the ID to the Location.

   ```
   ID,Location,Classification
   Data1,DB1,High
   ```

3) **cloudconfig.csv**

   Many cloud providers have services that provide inventories of how an organisation's cloud assets are configured to access each other. AWS config [71] is one such service. This information documents how data can move between cloud assets. In this example, we note which services (eg. database as a service) an asset (eg. container) has access to. In this case, we will have two-way dependencies between the asset and services. Additional information might allow the dependency to go one way. For example, an indication that the asset has read-only access would allow us to conclude that the dependency goes from the service to the asset but not the other way around.

   ```
   Asset,Service
   WWW1,DB1
   ```

4) **cryptoinventory.csv**

   Some services or software packages (e.g., Cryptosense [72]) can automatically generate an inventory of cryptographic keys and certificates present in machines, containers, and similar assets. Here we show some typical information available. In this example, there will be dependencies from the ID to the location and to the algorithm with the key size as a configuration parameter. Depending on the type of cryptographic asset, there may be other dependencies, such as on other keys or certificates, according to the rules of thumb.

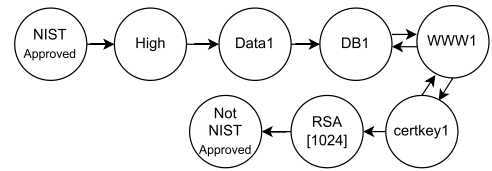

**FIGURE 10.** Security dependencies: Direct path from NIST approved to Not approved cryptographic object.

```
ID,Location,Type,Algorithm,Keysize
certkey1,WWW1,private key,RSA,1024
```

5) **crypto.json**

   Ideally, there would be a publicly available cryptographic registry which documents the dependencies and/or security levels of standard protocols and primitives. Here we have a minimal example in JSON format showing a single configuration with a single flag indicating the key size. There will be a dependency from each configuration to the corresponding security level. We have adopted the notation RSA[1024], indicating the primitive RSA with configuration flag 1024, for cryptographic protocols and primitives. The JSON object can list multiple ways of rating the security. Here, we have the security rated in bits, and whether the primitive is NIST approved. For our example, only the NIST-approved field is in use, and we will have a dependency from RSA[1024] to not-NIST-approved.

```
{
''name'': ''RSA'',
''configurations'':  [
{
''flags'': [ ''1024'' ],
''security'': 80,
''NIST-approval'':
''not-NIST-approved''
}
]
}
```

*b: ANALYSIS*

A reasonably simple program would be able to automatically generate the necessary dependencies for all entities by adding vertices and edges according to the rules discussed above for each row/object in each file. The complete graph is shown in FIGURE 10.

As can be seen from the graph, there is a directed path from NIST-approved to not-NIST-approved, indicating a situation where a data asset may not be adequately protected by the cryptography in use. It is straightforward to discover such directed paths using standard graph-theoretic algorithms.

Although this example is extremely simple, the same principles can be automated and applied to many larger data sets to discover potential security problems. Once the types of dependencies have been identified in the data sets (i.e. between columns), the analysis can be run automatically whenever information changes, or even to explore hypothetical 'what-if' scenarios for planning purposes, without any manual intervention.

## IX. CONCLUSION

The rapid development of quantum computing presents a substantial threat to the security of organisational information assets protected by established cryptosystems. Therefore, the timely transition to a quantum-safe state is of the utmost importance. This paper presents a comprehensive migration framework that offers guidance to organisations for smooth transition. This framework enables organisations to conduct a comprehensive identification and evaluation of their cryptographic assets. This is essential in prioritising migration tasks to minimise the risk of harm to organisational information assets. In addition, this paper presents a comprehensive method for conducting a security dependency analysis using a graph-theoretic approach. This enables the identification of situations in which the employed cryptosystems fall short of the desired level of security for protecting the targeted information asset during the intended security duration. The effectiveness of the dependency analysis procedure was demonstrated by examining three distinct case studies. This demonstrates that our approach is useful in discovering the potential security problems and can be automated and applied to many larger scenarios.

While our framework is designed for comprehensive PQC deployment, including both small and large enterprises as well as cloud-based entities, it does not extend to specific technology-based niches, such as smaller devices or the Internet of Things (IOT), presenting an opportunity for future research. However, the inherent flexibility of our framework, particularly in the use of security dependency analysis, enables it to effectively integrate with other well-established security protocols, such as NIST CSF, ETSI transition strategies, CARAF, and the NNCS Handbook, which are also discussed in our paper. This multidimensional strategy ensures a thorough examination of all aspects of information security management, including risk assessment.

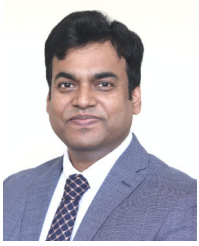

**KHONDOKAR FIDA HASAN** (Senior Member, IEEE) received the Ph.D. degree in computer science from the Queensland University of Technology (QUT), Australia. He has a long and distinguished academic career and a strong interest in cutting-edge research. He is currently a Lecturer in cyber security with the University of New South Wales (UNSW). He received a Higher Education Teaching Certification from Harvard University, USA. He has been titled a fellow of the Advanced Higher Education Academy, U.K., in recognition of his research and teaching excellence in Higher Academia. He is also a member of Australian Information Security Association (AISA). His current research interests include post quantum cryptography, enterprise security, and the application of artificial intelligence-based solution in smart cities and healthcare.

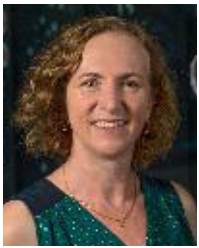

**LEONIE SIMPSON** received the Ph.D. degree from the Queensland University of Technology (QUT), Brisbane, QLD, Australia, in 2000. She has been involved in information security research for over 25 years. She has extensive experience analyzing cryptographic algorithms and finding weaknesses that reduce the security provided. She has applied her knowledge of design flaws in algorithms to help develop more secure ciphers, working in teams with Australian and international researchers. She is currently an Associate Professor and an Information Security Researcher with QUT. She is also researching efficient authenticated encryption methods for use in securing data transmissions between small and low-power devices in the rapidly growing the Internet of Things. Her main research interest includes symmetric cryptology, widely used for data protection. She is a member of the International Association for Cryptologic Research (IACR), the Australian Information Security Association (AISA), and the Australian Mathematical Society (AustMS).

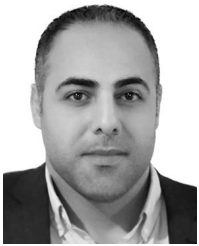

**MIR ALI REZAZADEH BAEE** (Senior Member, IEEE) received the Ph.D. degree in information security from the Queensland University of Technology (QUT), Brisbane, QLD, Australia. He has over 15 years of experience working in computer science, and since 2012, he has been involved in computer science research with a strong focus on applied cryptography and information security. His contributions have led to novel and important scientific outcomes. He is currently an Information Security Researcher with QUT, collaborating with the Cyber Security Cooperative Research Center (CSCRC), Australia, to solve pressing real-world cyber security challenges. He is a member of the International Association for Cryptologic Research (IACR). He has actively served as a Reviewer for flagship journals, such as the IEEE Transactions on Vehicular Technology, IEEE Transactions on Dependable and Secure Computing, IEEE Transactions on Intelligent Transportation systems, and IEEE Internet of Things Journal, and conferences, including the IACR's EUROCRYPT and ASIACRYPT.

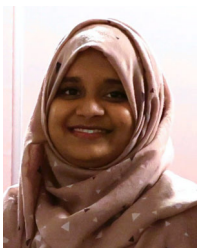

**CHADNI ISLAM** received the Ph.D. degree from the University of Adelaide, collaborating closely with the CSIRO's Data61 Distributed System Security Group, both renowned institutions in Australia. She is currently a Lecturer in cybersecurity with the School of Computer Science, Queensland University of Technology (QUT), Brisbane, Australia. Previously, she was a Postdoctoral Research Fellow with the CREST Research Group, School of Computer Science, The University of Adelaide, Australia. She possesses a deep passion for teaching and research, constantly seeking opportunities to expand her knowledge and delve into the realms of software engineering and cybersecurity. Her research interests include the practical application of natural language processing and machine learning techniques within the fields of software engineering and cybersecurity.

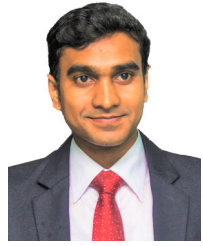

**ZIAUR RAHMAN** received the Ph.D. degree in cybersecurity from RMIT University, Melbourne, Australia. His Ph.D. thesis was selected as outstanding (top 5%). He is currently a Research Fellow in Cybersecurity for the Cyber Security Cooperative Research Council (CSCRC) Project with the Queensland University of Technology (QUT), Brisbane, Australia. Previously, he served academia overseas, government, industry, and internationally funded projects at Charles Sturt, Deakin, and RMIT University. He has published high quality publications in the top journals and conferences, including articles that were nominated and received best paper awards. He is affiliated with ACM, ACS, AISA, IEB, and IEEE. His research interests include blockchain technology, cryptography, the IoT security and privacy, edge machine learning, graph theory, and software engineering. He was a recipient of Higher Degree by Research (HDR) Award with RMIT University.

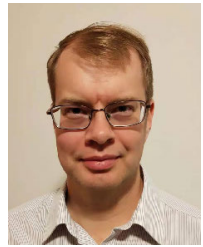

**WARREN ARMSTRONG** received the Ph.D. degree from The Australian National University, in 2011. He has 15 years of industry experience building cybersecurity products. He is currently the Director of Engineering with QuintessenceLabs Pty Ltd. He is the main point of engagement between QuintessenceLabs and Cyber Security Cooperative Research Centre.

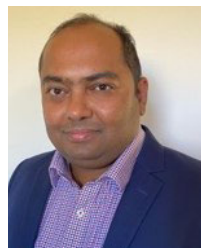

**PRAVEEN GAURAVARAM** received the Ph.D. degree. He is currently a Senior Scientist with Tata Consultancy Services Ltd. (TCS), Australia & New Zealand (ANZ). He primarily leads TCS's Cyber Security Research & Co-Innovation in Partnership with the Cyber Security Cooperative Research Centre (Cyber Security CRC) by actively engaging with the internal and external stakeholders to build, develop, and utilize emerging cyber security research. At TCS ANZ, he has also been influential in the business development, marketing, corporate affairs, and innovation branding activities. He is an Adjunct Associate Professor with the Faculty of Engineering, School of Computer Science and Engineering, University of New South Wales (UNSW), and an Adjunct Professor with the Faculty of Science, Engineering and Built Environment, School of Information Technology, Deakin University. He is also a member of the ICT Curriculum Advisory Group, Southern Cross University.

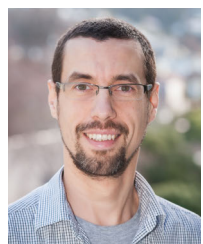

**MATTHEW MCKAGUE** received the B.Sc. degree (Hons.) in mathematics from the University of Regina, Regina, Canada, in 2004, and the M.Math. and Ph.D. degrees in combinatorics and optimization from the University of Waterloo, Waterloo, ON, Canada, in 2005 and 2010, respectively. He is currently a Lecturer in computer science with the Queensland University of Technology, Brisbane, Australia. Previously, he was a Research Fellow with the Centre for Quantum Technologies, Singapore, and a Lecturer with the Department of Computer Science, University of Otago, Dunedin, New Zealand.

• • •